# Stoichiometry Dependence of Potential Screening at $La_{(1-\delta)}Al_{(1+\delta)}O_3/SrTiO_3$ Interfaces

C. Weiland and J.C. Woicik

*National Institute of Standards and Technology, Gaithersburg, MD, USA*

G.E. Sterbinsky and A.K. Rumaiz

*National Synchrotron Light Source,*

*Brookhaven National Laboratory, Upton, NY, USA*

C.S. Hellberg

*Center for Computational Materials Science,*

*Naval Research Laboratory, Washington, DC, USA*

S. Zhu

*Department of Materials Science and Engineering,*

*Cornell University, Ithaca, NY, USA*

D.G. Schlom

*Department of Materials Science and Engineering,*

*Cornell University, Ithaca, NY, USA and*

*Kavli Institute for Nanoscale Science,*

*Cornell University, Ithaca, NY, USA*

ABSTRACT

Hard x-ray photoelectron spectroscopy (HAXPES) and variable kinetic energy x-ray photoelectron spectroscopy (VKE-XPS) analyses have been performed on 10 unit cell $La_{(1-\delta)}Al_{(1+\delta)}O_3$ films, with La:Al ratios of 1.1, 1.0, and 0.9, deposited on $SrTiO_3$. Of the three films, only the Al-rich film was known to have a conductive interface. VKE-XPS, coupled with maximum entropy analysis, shows significant differences in the compositional depth profile between the Al-rich, the La-rich, and stoichiometric films; significant La enrichment at the interface is observed in the La-rich and stoichiometric films, while the Al-rich shows little to no intermixing. Additionally, the La-rich and stoichiometric films show a high concentration of Al at the surface, which is not observed in the Al-rich film. HAXPES valence band (VB) analysis shows a broadening of the VB for the Al-rich sample relative to the stoichiometric and La-rich samples, which have insulating interfaces. This broadening is consistent with an electric field across the Al-rich film. These results are consistent with a defect-driven electronic reconstruction.

## I. INTRODUCTION

$LaAlO_3/SrTiO_3$ (LAO/STO) heterointerfaces have been a subject of intense study for the past few years [1–13]. Although both LAO and STO are band insulators, the interface between $TiO_2$-terminated STO and LAO is a high-mobility conductor when the thickness of the LAO layer is above a critical value [1–3]. Despite significant theoretical and experimental efforts, the origin of the interfacial conductivity remains a subject of intense debate. One of the most commonly-accepted scenarios to explain the interface conductance involves charge transfer between polar LAO and non-polar STO. The proposed charge transfer is a direct consequence of the polar discontinuity going from a non-polar STO layer to a polar LAO layer, which would otherwise give rise to a diverging electrostatic potential in the LAO layer. The electron transfer from the LAO to the STO layer is confined in the first one or two unit cells of the STO, forming a conductive two dimensional electron gas (2DEG) [1–4].

Alternately, some reports have attributed the observed conductance to oxygen vacancies in the STO [14, 15]. The electrons liberated by oxygen vacancies generated during the synthesis can populate the otherwise empty Ti $3d$ band in the STO giving rise to conductivity. However, studies have shown conductive interfaces remain after oxygen annealing to reduce oxygen defects, and thus the conductivity observed cannot be attributed to oxygen vacancies alone [16]. Additionally, the dependence of the interfacial conductivity on the termination of the STO cannot be explained in the context of oxygen vacancies.

2An additional explanation involves ionic interdiffusion or chemical reconstruction at the LAO/STO interface [7, 8], similar to that which has been seen to resolve polar discontinuities in semiconductor heterojunctions such as GaAs on Si or Ge [17]. La-doped STO is known to be conductive [18]; thus diffusion of La into the STO layer could lead to conductivity at the interface as well.

Recently, the stoichiometry of the LAO layer has been reported to greatly affect the interface properties [4, 19]. A recent study that combined experiments and first principles calculations revealed that a La:Al ratio $\leq 0.97 \pm 0.03$ is required for the formation of the 2DEG [4]. To demonstrate that non-stoichiometry is required for formation of a 2DEG, LAO films of various stoichiometries were deposited simultaneously by molecular beam epitaxy (MBE); the distance of the substrate from the La and Al sources in the chamber determined the film composition. Interfacial conductivity occurred only when the LAO layer was Al-rich (La:Al $\leq 0.97 \pm 0.03$). Because films were all grown simultaneously at the same temperature and partial pressure of oxygen, stoichiometry is the only parameter that varies between the films.

In this paper, we have used hard x-ray photoelectron spectroscopy (HAXPES) and variable kinetic energy x-ray photoelectron spectroscopy (VKE-XPS) to study the chemical and electronic structure of LAO/STO interfaces with La:Al ratios of 0.9, 1.0, and 1.1. The samples are 10 unit cell thick films of LAO deposited on TiO2-terminated single crystal STO substrates by MBE. Samples were deposited with a background of distilled ozone (partial pressure $1 \times 10^{-6}$ Torr). The LAO stoichiometry was dependent on the location of the substrate with respect to the source. A detailed description of the sample deposition procedure and characterization are presented elsewhere [4]. No further sample treatments were applied prior to analysis.

## II. METHODS

### A. HAXPES Measurements

HAXPES measurements were performed at the National Institute of Standards and Technology HAXPES facility at beamline X24A of the National Synchrotron Light Source at Brookhaven National Laboratory. The beamline has an energy range of 2.1 to 6 keV using a double Si (111) crystal monochromator. Spectra are recorded using a 10 keV 400 mm diameter hemispherical electron energy analyzer, with the detector aperture set to a 300 $\mu$m and mounted at 90° with respect to the beam axis. Core-level spectra were recorded at 200 eV pass energy and valence-band (VB) spectra at 500 eV pass energy. Samples were oriented at an 85° photoelectron takeoff angle, defined between sample surface and detector axis. Pressure in the analysis chamber was maintained below $10^{-8}$ Torr. The beam spot was defined by 0.5 mm vertical and 1 mm horizontal apertures. Spectra were recorded in an angular lens mode to ensure consistent relative peak intensities [20].

The energy scales for all samples were referenced to a Sr $3d_{5/2}$ signal collected concurrently with the core level under investigation. To eliminate experimental charging, a small amount of conductive silver paste was placed on the corners of the samples to ensure a conductive path from the surface to the grounded sample holder. Sample positions were aligned to ensure no photoemission signal from the silver paste was observed. Additionally, each individual cycle, consisting of one to five scans of the core level under investigation and a single scan of the Sr $3d$, was saved independently and summed for the spectra presented here. The relative energy shifts between the Sr $3d_{5/2}$ and the core level under investigation were checked between each cycle to further ensure no energy shift due to differential charging was present.

### B. Theoretical Calculations

Density functional theory (DFT) calculations used the generalized gradient approximation as implemented in VASP [21–23]. Projector-augmented wave functions with a 282.9 eV plane wave cutoff were used [24]. The in-plane lattice constant was fixed to the theoretical SrTiO$_3$ lattice constant, $a_{SrTiO3}$ = 3.948 Å. For partial density of states calculations of stoichiometric LaAlO$_3$, a 5-atom cell and $32 \times 32 \times 32$ Monkhorst-Pack $k$-points were used; the bulk defect calculations used larger cells and correspondingly reduced $k$-point grids. Calculations were performed for La$_{(1-\delta)}$Al$_{(1+\delta)}$O$_3$ and contain no electric fields.

For the calculations of stable surface structures, the equivalent of $6 \times 6$ Monkhorst-Pack $k$-points for a 1×1 surface cell were used. Geometries were relaxed until the residual forces were less than 0.01 eV/Å. At least 11 LaAlO$_3$ bilayers were included in the slabs, and the slabs are separated in the $z$-direction by at least 8 Å of vacuum. The bottom layer is AlM$_2$, where M is a virtual atom with atomic number 8.25 (between O and F). The AlM$_2$ layer has nominal charge -0.5|$e$|, creating an insulating bottom surface with no electric field in the LaAlO$_3$ [25, 26]. We have verified that a slab terminated on both top and bottom by AlM$_2$ has no mid-gap states.

The surface energy is given by:

$$G = E - \mu_{La}N_{La} - \mu_{Al}N_{Al} - \mu_O N_O \qquad (1)$$

where $E$ is the total energy, $N_i$ is the number of each atom in the calculation, and $\mu_i$ is the chemical potential of each species. The potentials are subject to the constraint:

$$\mu_{La} + \mu_{Al} + 3\mu_O = \mu_{LaAlO3}^{bulk}, \qquad (2)$$

where $\mu_{LaAlO3}^{bulk}$ is the computed energy of bulk LaAlO$_3$ strained to the SrTiO$_3$ lattice constant and relaxed in the $z$-direction. Growth occurs at a constant oxygen pressure, $10^{-6}$ Torr at 680°C, which fixes the oxygen chemical potential $\mu_O = -1.9$ eV [4, 27]. The anions, on the other hand, stick to the growing surface and are not in equilibrium with a reservoir at fixed temperature and pressure. As shown below, the anion chemical potentials will vary with the stoichiometry of the deposited material and the thickness of the film in a discontinuous way. We used $\mu_{La}$ as the independent parameter and determine $\mu_{Al}$ from Eq. (2).

### C. Maximum Entropy Calculations

Calculations of depth profiles from VKE-XPS data were performed using a maximum entropy regularization method (MEM). A detailed exposition of the MEM regularization applied to VKE-XPS data can be found elsewhere [28]. Briefly, the determination of a concentration versus depth profile from VKE, or more commonly angle-resolved XPS is an ill-conditioned problem, requiring a regularization function to avoid over-fitting (fitting the noise). This can be accomplished by maximizing a functional Q:

$$Q = \alpha S - \frac{\chi^2}{2}. \qquad (3)$$

Here, $S$ is the regularization function, which is always negative, and $\chi^2$ is calculated between the measured and calculated VKE-XPS data. The regularization parameter, $\alpha$, is varied to provide the optimized spectra which is the best fit (minimum $\chi^2$) and most regularized (maximum $S$). For the regularization function, $S$, a maximum entropy function was used:

$$S = \sum_i \sum_j n_{i,j} - m_{i,j} - n_{i,j} \log\left(\frac{n_{i,j}}{m_{i,j}}\right). \quad (4)$$

The function $S$ compares the concentration $n_{i,j}$ of a species $j$ as a function of depth '$i$' with that for an initial model $m_{i,j}$. The quality of fit for this $S$ is dependent on the quality of the initial model [28], but here the thickness of the LAO films are known [4]. For the initial model, a 3.79 nm thick film with 1:1 La:Al was used for all samples.

Once a model depth profile is created, the corresponding VKE-XPS spectra are calculated according to the Beer-Lambert law. The sample is modeled as a series of thin slabs, with in slab intensity:

$$I_{j,k} = R_{j,k}\left[1 - \exp\left(-\frac{\Delta t}{\lambda_{j,k} \sin \theta_k}\right)\right] \sum_{i=0}^N n_{i,j} \exp\left(-\frac{t_i}{\lambda_{j,k} \sin \theta_k}\right). \quad (5)$$

Here $R_{j,k}$ is the photoemission from a pure sample of species $j$ at photon energy $k$, and the thickness of each slab is given by $\Delta t$. The inelastic mean free path of the photoelectron $\lambda_{j,k}$ is calculated form the Tanuma, Powell, and Penn (TPP-2M) equation [29]. A takeoff angle, $\theta$, of 85° was used for all beam energies to match the experimental data. Since the photoemission intensity of a pure sample is practically unknown and depends on experimental factors that may change with time, such as the x-ray flux, sp ot size, and detector efficiency, it is more practical to consider the ratio $I/R$. Using this ratio, an apparent atomic concentration can be calculated as:

$$\frac{\frac{I_{j,k}}{R_{j,k}}}{\sum_j \frac{I_{j,k}}{R_{j,k}}}. \quad (6)$$

This value can be compared to the measured data to provide $\chi^2$. The model $n_{i,j}$ values are then iterated until the functional $Q$ is maximized.

### III. RESULTS AND DISCUSSION

10 unit-cell thick La-rich, Al-rich, and stoichiometric LAO, with La:Al ratios of 1.1, 0.9, and 1.0 respectively, were analyzed with HAXPES. Of the three films, only the Alrich produced a conductive interface as determined by resistivity and low temperature Hall measurements [4]. Sr $3d$, La $3d_{5/2}$, Al $1s$, and Ti $2p$ core levels collected at $h\nu = 2150$ eV are displayed in Figure 1. Within the sensitivity of the measurement, no appreciable differences are observed between the three samples, and all elements are observed to be in a single chemical state (the Ti $2p$ and Sr $3d$ show spin orbit doublet, while the La $3d_{5/2}$ exhibits an energy loss satellite feature). However, dopant-level intermixing may be below the HAXPES sensitivity which is typically on the order of one atomic percent, and thus not observable in these measurements.

Close inspection of the Ti $2p$ feature finds no evidence of the Ti$^{3+}$ binding state, which would be centered about 2 eV higher in kinetic energy from the main Ti$^{4+}$ feature. This is true even for the Al-rich sample. The Ti$^{3+}$ feature has frequently been used as evidence of electronic reconstruction of the interface, as the Ti $3d$ becomes populated with the transferred electron [10]. However, the presence of Ti$^{3+}$ may also be evidence of chemical reconstruction [8] or O vacancies [30]. The absence of a Ti$^{3+}$ signal here may be related to the measurement sensitivity for the relatively bulk-sensitive HAXPES measurement.

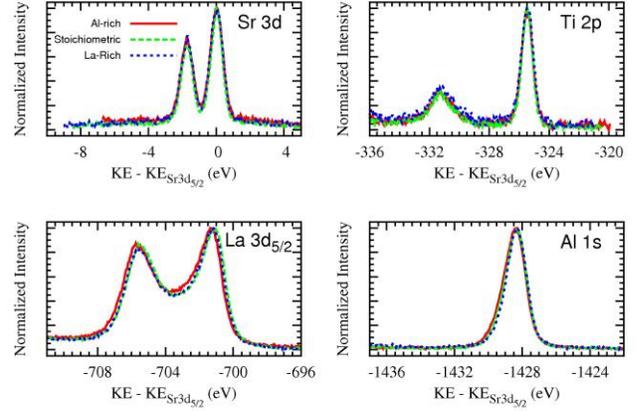

FIG. 1. Core-level spectra from Al-rich, stoichiometric, and La-rich LAO/STO films collected with photon energy of 2150 eV. Kinetic energies have been referenced to the Sr $3d_{5/2}$ core level, and intensities normalized to the maximum value.

VB spectra of the three films collected at $h\nu = 2150$ eV are shown in Figure 2. Spectra from all three films show similar features: a large, central feature constructed primarily of La $5p$-like states, a shoulder on the low kinetic energy side dominated by Al $3s$ states, and a shoulder on the high kinetic energy side dominated by O $2p$ states. The spectra are consistent with previous measurements of LAO and show no obvious signal from the underlying STO substrate [31].

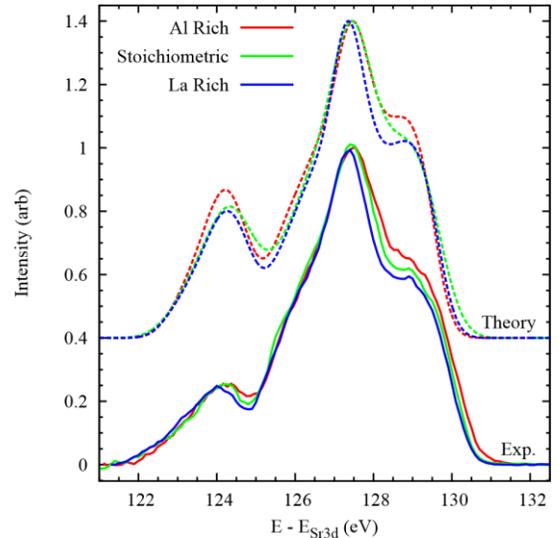

FIG. 2. VB spectra from Al-rich, stoichiometric, and La-rich LAO/STO films collected with photon energy of $h\nu = 2150$ eV. Kinetic energies have been referenced to the Sr $3d_{5/2}$ core level and intensities normalized to the maximum value

Also shown in Figure 2 are the theoretical HAXPES VB spectra reconstructed from *ab initio* partial density of states calculations, following Ref. [32]. The theoretical VB spectra are aligned with the maximum intensity feature in the corresponding experimental spectrum. Since photoelectron spectroscopy does not directly probe the density of states, but instead measures transition probabilities between the initial ground state to excited final state [33], the experimental VB is not simply the calculated total DOS but rather the density of states modulated by the electronic transition probability as given by [32]:

$$I(E, h\nu) \propto \sum_{i,l} \rho_{i,l}(E) \sigma_{i,l}(E, h\nu). \qquad (7)$$

Here $E$ is the photoelectron binding energy, $h\nu$ is the incident photon energy, $\rho_{i,l}(E)$ is the angular momentum $l$-resolved partial density of states of the $i^{th}$ atom, and $\sigma_{i,l}(E,h\nu)$ is the photoionization cross section. Photoionization cross sections were taken from Refs. [34 and 35]. The data is further convolved with a Gaussian function (sigma=1) to account for experimental broadening. Excellent agreement is observed between theoretical and experimental VB structures.

A +0.25 eV kinetic energy shift of the VB edge is observed for the Al-rich sample with respect to the other two films. Close inspection of the VB shape suggests that the shift results from a broadening of the VB features and not a distinct state. This is explicitly evident from the comparison of experimental and theoretical spectra: the VB edges of the calculated Al and La rich samples are seen to be roughly equivalent, yet in the experimental data, the Al-rich VB edge is clearly shifted to higher kinetic energy. This VB broadening is consistent with an electric field across the LAO film, as discussed further below.

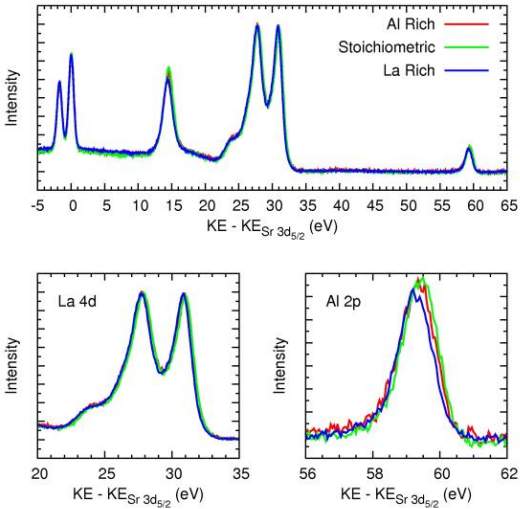

FIG. 3. Shallow core levels for Al-rich, stoichiometric, and La-rich LAO/STO films collected at photon energy of 3500 eV. Kinetic energies are referenced to the Sr $3d_{5/2}$ core level and intensities are normalized to the Sr $3d_{5/2}$ intensity.

VKE-XPS was used to probe the sharpness of the interface. VKE-XPS uses HAXPES spectra collected at various photon energies to tune the depth sensitivity of the measurement. The Sr $3d$, La $4d$, and Al $2p$ features were selected for VKE-XPS analysis as they all fall within a 60 eV range in binding energy, and therefore have similar attenuation lengths. An example of the energy window collected at $h\nu$ = 3500 eV is shown in Figure 3. As with the La $3d_{5/2}$ (Fig. 1), no significant differences are seen in the La $4d$ for the three samples. Some subtle peak intensity and position shifts are observed, however, for the Al $2p$. Similar intensities are observed for the Al-rich and stoichiometric samples, and a lower intensity for the La-rich sample. Considering the roughly equivalent intensities observed for the La $4d$ and the noise level due to the low cross section of the Al $2p$, the intensity trend is consistent with the expected La:Al ratios. The Al $2p$ for the La-rich LAO is also found to be shifted approximately 0.1 eV to higher binding energy; this shift is attributed to the different defects present in the films necessary to accommodate off-stoichiometric conditions. Note that when the Al $2p$ peak is aligned and normalized in intensity (not shown), no broadening is apparent, in contrast to the VB.

VKE-XPS trends are shown in Figure 4. Intensities (I) are plotted as the ratio:

$$\frac{I_M}{(I_{La4d_{5/2}} + I_{Al2p_{3/2}} + I_{Sr3d_{5/2}})}, \qquad (8)$$

where M = La$4d_{5/2}$, Al$2p_{3/2}$, or Sr$3d_{5/2}$, and the peak intensities have been normalized to the photoionization cross sections so that the ratios only reflect the depth-dependent concentration. The trends observed in Figure 4 show distinct differences between the samples. For the Al-rich sample, the La and Al ratios track each other quite closely. This demonstrates a homogeneous La:Al ratio throughout the film. The La-rich film, as expected, shows a higher La signal relative to the Al signal. However, the La and Al ratios diverge with beam energy. This photon energy dependence is also observed for the stoichiometric sample. In both cases, the La ratio increases with respect to the Al at the highest beam energies, which correspond to the greatest analysis depths. This demonstrates that the La:Al ratio is not homogeneous throughout the La-rich and stoichiometric films, with higher La concentration towards the interface. This is consistent with B-site vacancies near the interface, as recently proposed [4].

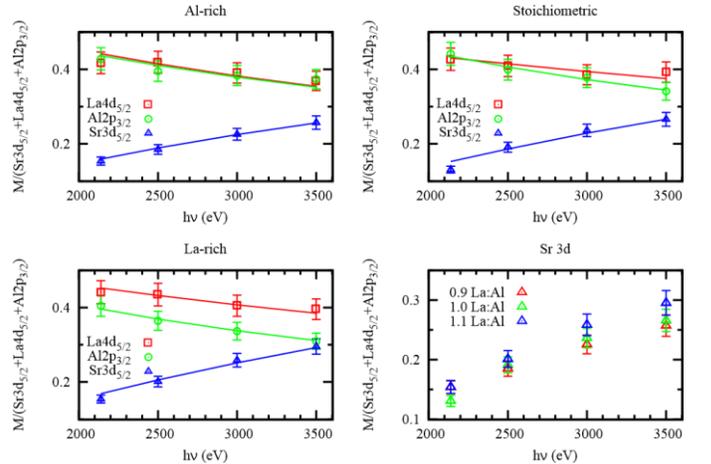

FIG. 4. VKE-XPS results showing intensities plotted as a function of photon energy (symbols). Peak areas are normalized to the photoionization cross section and error bars are estimated from the signal-to-noise ratio. Also shown is an overlay of the Sr $3d_{5/2}$ ratio for the three samples. Solid lines show the calculated VKE-XPS data from the results of a maximum entropy calculation. Excellent agreement is achieved between the calculated and experimental results.

Also of interest is the Sr $3d_{5/2}$ signal. The three films were deposited simultaneously; thus the photon energy dependence of the Sr $3d_{5/2}$ signal should be roughly the same across all films. As seen in the overlay of the Sr $3d_{5/2}$ ratios for the three samples, also shown in Figure 4, the signal is clearly different for each sample. Because of the different compositions and vacancy densities expected between the three films, there may be some difference in electron attenuation; however, calculations of the IMFP suggest that the largest variation between samples will be less than 0.2 Å for a given photon energy. The observed differences between the samples thus cannot be explained by the different attenuation alone, and must reflect some difference in the compositional profiles of the samples.

Further confirmation of different compositional profiles between LAO films is found comparing the measured VKE-XPS data to simulated spectra. Spectra were simulated using the NIST database for the Simulation of Electron Spectra for Surface Analysis (SESSA) [36]. The SESSA database contains data for a number of parameters for the calculation of simulated HAXPES spectra, including the photoionization cross-sections, IMFPs, and effective attenuation lengths. In the simulations, the LAO films were modeled as 20 half unit cell layers, while the STO substrate was modeled as a single homogeneous substrate. Compositions of the LAO layers were determined from first principles calculations. The cation vacancy density, $v$, at the Al-rich interface depends on diffusion barriers and cannot be determined from our first principles calculations, so multiple values for $v$ were simulated. For the simulations, $v = 0$ and 0.167 were found to show no significant differences due to the strong attenuation from the interfacial layer. For the STO substrate, an atomic density of $8.397 \times 10^{22}$ $cm^{-3}$ and band gap of 3.25 eV were used. Each half-layer in the LAO film was given an atomic density of $9.18 \times 10^{22}$ $cm^{-3}$, a band gap of 6.2 eV, and thickness of 1.9525 Å, or half the lattice constant. The Al $2p_{1/2}$, La $4d_{5/2}$, and Sr $3d_{5/2}$ peak intensities were modeled as functions of photon energy, $h\nu = 2150$, 2500, 3000, and 3500 eV to match the measured data.

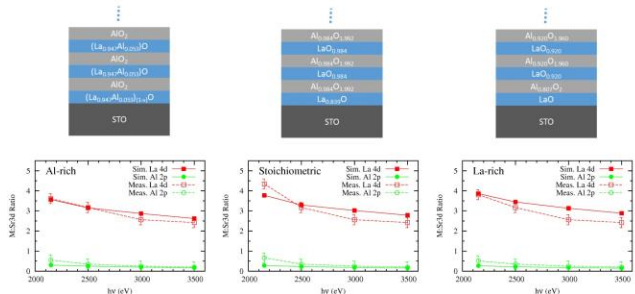

FIG. 5. La:Sr and Al:Sr ratios as a function of photon energy for LAO films on STO, simulated using SESSA (closed symbols) and measured with HAXPES (open symbols). Model chemical structures used for the simulations are shown above each plot. For the Al-rich sample, interfacial vacancy densities, $v$, of 0 and 0.167 were simulated and showed no significant differences; $v = 0.167$ is shown in the plot. Error bars in measured data are estimated from the observed signal to noise in the measurements.

The simulated La:Sr and Al:Sr ratios are shown in Figure 5, overlayed with the raw measured ratios (i.e. without normalizing to the photoionization cross sections). For the Al-rich sample, the simulated La:Sr ratio tracks very closely the experimental data. For all samples, the calculated Al:Sr ratio is lower than the measured one; this could be due to a higher Al concentration, especially at the surface. In the stoichiometric and La-rich films, the simulated La:Sr ratios are very different compared with the measured data. The La profiles for the stoichiometric and La-rich films are not consistent with homogeneous La distribution throughout the LAO.

To probe the concentration versus depth profile in the three samples, a maximum entropy method (MEM) algorithm was executed. Results are shown in Figure 6. Calculated VKEXPS data from the MEM-produced depth profiles match the experimental data quite closely (Figure 4), with a maximum RMS deviation of 1%. As expected from the VKE-XPS data, the resulting depth profiles show significant differences between samples. For the Al-rich sample, the La and Al signals track closely and quickly decay in the substrate region; a relatively abrupt interface is observed, compared to the La-rich and stoichiometric samples. The profiles for the two insulating samples are quite similar, showing an Al-rich surface and significant La diffusion into the bulk. It must be noted that the error in concentration increases with depth [28], due to the attenuation of the signal; thus the concentration of the La diffusion cannot be precisely quantified at these depths. Error bars in the figure are estimated by the total change for each individual data point that would lead to the measured uncertainty. While this is likely an overestimation of the error as all all uncertainty is borne by each individual data point, it does demonstrate the rapid increase in uncertainty with increasing depth. However, it is clear that some La enrichment at the interface must occur to produce the observed VKE-XPS data. The Al profiles for the La-rich and stoichiometric samples do not show any obvious diffusion into the STO substrate. STEM-EELS measurements on similarly fabricated samples showed subtle intermixing at the interface for both La-rich and Al-rich samples [4], and indeed some intermixing may be present in the Al-rich MEM profile, but it is difficult to compare STEM-EELS and MEM results due to the high uncertainty at the interface region for MEM and the global nature of MEM as compared to the local nature of STEM/EELS.

Four main points can be drawn from these results: (1) interfacial La enrichment is observed in the insulating La-rich and stoichiometric samples, but not in the conductive Al-rich sample, (2) An Al rich surface is observed in the La-rich and stoichiometric films, but not in the Al-rich film, (3) no chemical shifts or multiple binding states are observed in either the STO substrates or LAO films, and (4) the conductive, Al-rich film shows VB broadening. The observed Al enrichment on the surface of the La-rich and stoichiometric samples can be understood using Density Functional Theory (DFT) calculations. The surface energies of 104 candidate surfaces of LaAlO$_3$ strained to the SrTiO$_3$ lattice constant were calculated. The bulk of the LaAlO$_3$ was stoichiometric for the calculations, but the surface composition was varied. The energies of the low-energy (001) surface structures are shown in Fig. 7.

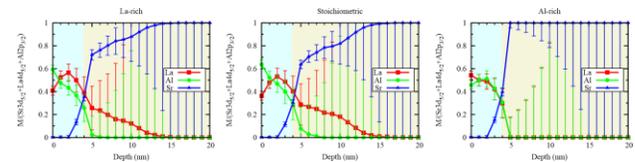

FIG. 6. Maximum entropy depth profiles for La-rich, stoichiometric, and Al-rich LAO on STO. Error bars are estimated from the total change per data point which would result in the observed measurement uncertainty.

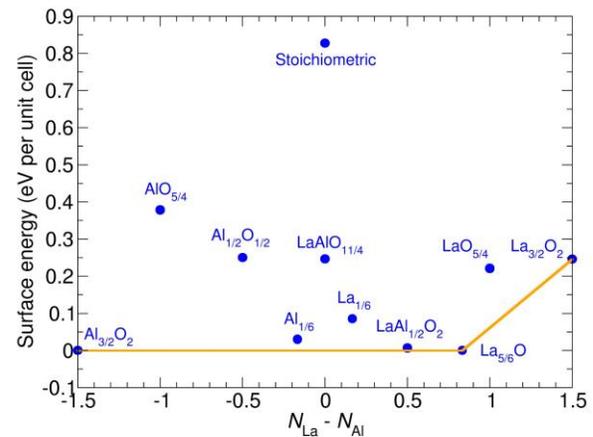

FIG. 7. Computed surface energies of relevant structures as a function of cation concentration per surface unit cell. Structures are labelled by their surface layers. The outermost "bulk" layer in each case is AlO$_2$; the "Stoichiometric" surface is terminated by an ideal AlO$_2$ layer. The surface has surprisingly few ground states, just Al$_{3/2}$O$_2$, La$_{5/6}$O, and La$_{3/2}$O$_2$; the other states can reduce their energies to the orange tie lines by phase separating. In particular, no stoichiometric ($N_{La} = N_{Al}$) surface is stable.

As the bottom of the film bonded to the STO is a LaO layer, the stoichiometric surface, labeled "Stoichiometric" in Fig. 7, is terminated by an AlO$_2$ layer. We label the other surfaces by their outermost layers, where the first complete layer of AlO$_2$ is defined as the beginning of the bulk. The lowest energy surface with stoichiometric cation concentrations is LaAlO$_{11/4}$, which is equivalent to the stoichiometric LaAlO$_3$ surface with with 1/4 monolayer of oxygen vacancies[37]. However, the LaAlO$_{11/4}$ surface is unstable to phase separation. Of the surfaces with stoichiometry $|N_{La} - N_{Al}| \leq 1.5$, only three are globally stable: Al$_{3/2}$O$_2$ is the only Al-rich ground state, while La$_{5/6}$O and La$_{3/2}$O$_2$ are La-rich ground states. The Al$_{3/2}$O$_2$ and La$_{5/6}$O surfaces are shown in Figure 8.

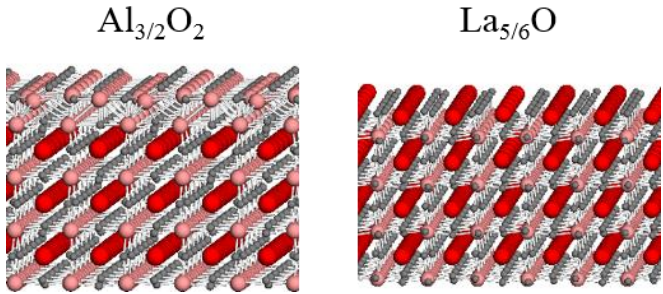

FIG. 8. Globally stable LAO surfaces: Al-rich Al$_{3/2}$O$_2$ and La-rich La$_{5/6}$O. Other surfaces will phase separate into these two surfaces. La atoms are red, Al are pink, and O are gray.

If we ignore for now defects at the interface and in the bulk of the film, stoichiometric growth will initially phase separate into Al$_{1/6}$ and La$_{1/6}$ regions, each covering half the surface. With further cation surface diffusion, the surface can phase separate into a sequence of surface phases with cation concentrations $N_{La} - N_{Al}$ further from zero, ultimately ending with Al$_{3/2}$O$_2$ and La$_{5/6}$O regions on the surface, which have the lowest surface energies.

La interdiffusion is a potential source of conductivity at LAO/STO interfaces, as La doped STO is known to have high mobility [18]. If La diffuses into the STO, then one would expect a conductive interface to occur. However, the VKE-XPS results presented here show that the samples with insulating interfaces have La enrichment at the interface, sample with conductive interface does not. While this could suggest La diffusion for the stoichiometric and La-rich samples, the MEM uncertainty at those depths precludes definitive characterization.

While a low level of La diffusion cannot be ruled out for the Al-rich interface, the different surfaces observed (Figure 6) point to a more complex phenomenon. Notably, the La-rich and stoichiometric films exhibit slightly Al-rich surfaces, while the Al-rich sample shows a slightly La-rich surface. This seemingly puzzling result can be understood by considering the various stable surfaces along with the defects that form in the bulk of the films. The chemical potentials will be equal throughout the film if the kinetic barriers are small relative to the growth temperature. Equal chemical potentials will still likely result in different stoichiometries at the interface, the surface, and in the bulk of the film. Additionally, significant kinetic barriers are to be expected to create or fill in the extended Al$_2$O$_3$ vacancy complexes which form during La-rich or stoichiometric growth.[4] Thus the density of vacancy complexes may become fixed during the initial growth of the film. In particular, if the initial vacancy complex density results in a film stoichiometry with a La/Al ratio that is higher than the La/Al ratio of the growth flux, Al will accumulate on the surface, forming an increasing fraction of the Al$_{3/2}$O$_2$ surface as the film grows thicker.

For Al-rich growth, the Al$_{La}$ substitutional defects that form are point defects whose density can vary throughout the bulk of the film. Thus, in this case, the surface structure can remain constant during growth, with no Al or La accumulation on the surface.

In addition to the surface structures, cation vacancies form at the interface[4]. For the stoichiometric and La-rich films, the vacancies are free to diffuse through the Al$_2$O$_3$ vacancy complexes in the film. The vacancies screen the charge from the polar LAO film, leading to the absence of electric field in the sample. However, in the Al-rich film, which only has point defects, the diffusion barrier is higher, restricting the total amount of defects that may form at the interface. The charge is thus only partially screened and a small electric field is present in the film. This electric field is observed in the HAXPES VB measurement (Figure 2).

To demonstrate the electric-field induced broadening, the VB for the Al-rich LAO sample was modeled as a sum of a series of 10 unit cell LAO VB spectra. The VB spectrum for each individual unit cell was taken as an average of the VB spectra for the stoichiometric and La-rich unit cells. A linear shift of "$\varphi$" eV per unit cell and a constant offset of "$\zeta$" eV were then applied to the model unit cells. The unit cell spectra were attenuated according to their depth, $t$, using the Beer-Lambert law. The IMFP, $\lambda$, was taken as 3.4 nm, calculated from the TPP-2M equation [29]. The experimental takeoff angle, $\theta$, of 85° was used and the depth was taken as the mid-point of the unit cell. The resulting sum of the unit cell spectra was then normalized and compared with the Al-rich experimental data. The residual between the model and experimental spectra was minimized using a least squares optimization of $\varphi$ and $\zeta$. This model is based on the assumption of flat bands for the La-rich and stoichiometric LAO films; however, even if there exists a potential or band bending across the La-rich and stoichiometric films, the model will show the additional potential in the Al-rich film. The resulting slope was found to be 0.09 eV/unit cell (34 meV/°A), and the offset -0.39 eV (with respect to the La-rich and stoichiometric samples). This electric field is in close agreement with prior measurements using cross-sectional STM [13]. Singh-Bhalla, *et al.*, however, observed a larger electric field, of 80.1 meV/°A[38]. In that work, a metal electrode was deposited on the surface of the LAO layer, which may, as noted by the authors, affect the resulting electric field. The resulting band diagram, along with the modeled VB is shown in Figure 9a and b.

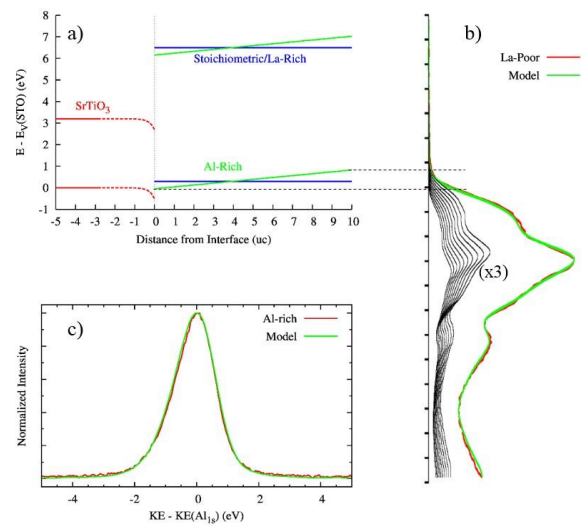

FIG. 9. (a) Band diagram for LAO/STO samples based on linear shift of attenuated unit cell VB spectrum. (b) VB spectrum showing the modeled unit cell curves. (c) The Al 1$s$ curve overlayed with the modeled broadened curve.

Significant core-level broadening was not observed here or in prior research [12]; however VB spectra were not reported in those works and the authors attempted to fit their data with a much larger electric field than measured here. Given the relatively small electric field, 0.09 eV per unit cell, determined from the VB, together with the relatively higher surface sensitivity of the core levels, the broadening in the core levels is expected to be minimal. For the Al 1$s$, for example, the full-width at half max (FWHM) of the core line is expected to broaden from 1.6 eV (for the stoichiometric and La-rich samples), to 1.7 eV, for the Al-rich sample (Figure 4c). In practice such a small broadening is difficult to confirm given the total experimental and intrinsic resolution in photoelectron spectroscopy. Our results are consistent with the recently-proposed model of stoichiometry dependence of LAO/STO interface conductivity [4]. In the stoichiometric and La-rich films, the polar catastrophe is averted by charge screening from cation vacancies at the interface; however, for the Al-rich film the aluminum substitutional defects block cation migration leaving only a partially screened interface. This allows a small residual electric field in the LAO layer; after a critical thickness is reached, a surface reconstruction becomes energetically favorable, transferring charge to the interface and resulting in the observed interface conductivity

### IV. CONCLUSIONS

In summary, we have used HAXPES and VKE-XPS to probe the LAO/STO interface for La-rich (1.1 La:Al), stoichiometric (1.0 La:Al), and Al-rich (0.9 La:Al) LAO. A small electric field of 34 meV/°A was observed from broadening of the VB in the Al-rich LAO with conductive interface, but no electric field was found in the stoichiometric and La-rich samples which had insulating interfaces. Additionally VKE-XPS demonstrated a sharper interface for the Al-rich LAO film, while the La-rich and stoichiometric films showed a greater level of interfacial intermixing. MEM-reconstructed depth profiles suggest Al enrichment on the surface and La enrichment at the interface for the stoichiometric and La-rich films, but not the Al-rich film. These results are consistent with an off-stoichiometry defect-driven electronic reconstruction model of conductivity at LAO/STO interfaces.

### V. ACKNOWLEDGEMENTS

Use of the National Synchrotron Light Source, Brookhaven National Laboratory, was supported by the U.S. Department of Energy, Office of Science, Office of Basic Energy Sciences, under Contract No. DE-AC02-98CH10886. The authors would like to thank Dr. Scott Chambers at Pacific Northwest National Laboratory for useful discussions. C. H. acknowledges support from the Office of Naval Research through the Naval Research Laboratory's Basic Research Program. Computations were performed at the AFRL and ERDC DoD Major Shared Resource Centers